# Molecular Beam Epitaxy of $Mn_2In_2Se_5$ van der Waals Layers Using Mn Intercalation

*Qihua Zhang[1*], Ke Wang[2], Wesley Auker[2], Maria Hilse[1,2,3], Stephanie Law[1,2,3,4]**

[1]Two Dimensional Crystal Consortium Materials Innovation Platform, The Pennsylvania State University, University Park, PA 16802, USA

[2]Materials Research Institute, The Pennsylvania State University, University Park, Pennsylvania 16802 USA

[3]Department of Materials Science and Engineering, The Pennsylvania State University, University Park, Pennsylvania 16802 USA

[4]Institute of Energy and the Environment, The Pennsylvania State University, University Park, PA 16802 USA

ABSTRACT The weak van der Waals (vdW) force in layered chalcogenide materials has enabled the growth of ternary chalcogenide layers using unconventional approaches. Here, we report the molecular beam epitaxy (MBE) growth of $Mn_2In_2Se_5$, a spin glass material with high level of magnetic frustration, through the heterointegration of MnSe on $In_2Se_3$. Directly depositing α-MnSe on the vdW $In_2Se_3$ layers results in Mn intercalation, transforming the $In_2Se_3$ layer into $Mn_2In_2Se_5$. Large growth windows, including substrate temperatures from 250-450 °C and Se:Mn flux ratio of 1.1-3.1, have been identified for the intercalation process. With an optimized MnSe deposition time, smooth, single-crystalline, and (0001)-oriented $Mn_2In_2Se_5$ layers with a root-mean-square (RMS) roughness of 1.5 nm can be synthesized. Further extending the MnSe deposition time results in the growth of uniform rock-salt structured α-MnSe (111) layers with a thickness of up to 8 nm and a narrow full-width-at-half-maximum (FWHM) of 0.35° in MnSe (222) XRD rocking curves. This report presents a unique approach for the growths of uniform and single-crystalline $Mn_2In_2Se_5$ vdW layers using MBE, and potentially opens a pathway for synthesis of ternary vdW chalcogenides by intercalation of new atomic species in binary vdW chalcogenides.

## INTRODUCTION

Two-dimensional (2D) materials have a range of unique material properties and broad device applications including low-power electronics, spintronics, and optoelectronics.[1-6] Layered chalcogenides are a subset of 2D materials characterized by their van der Waals (vdW) layered structure, where atoms within each layer (intralayer) exhibit strong covalent bonding, while each layer is bonded to the next by a weak vdW force. The weak interlayer vdW bonding in the layered chalcogenides results in variations of crystal structures (crystal symmetry, atomic stacking sequence, etc.) among these materials, which in turn leads to a wide range of unique electrical and optical properties that depend on their thicknesses, polytypes, and phases.[7-11]

In addition to synthesizing binary chalcogenides, we can combine two binaries to create a ternary chalcogenide. Ternaries often possess unique properties: for example, $MnBi_2Te_4$ is a ternary chalcogenide that displays both the properties of $Bi_2Te_3$ (a vdW topological insulator)[12, 13] and MnTe (an antiferromagnet)[14-16]. The resulting magnetic topological insulator has properties distinct from either of the binary end members.[17-20] The relatively weak vdW bonding between the layers in the vdW materials means that one can synthesize $MnBi_2Te_4$ thin films using nonconventional approaches. Generally, $MnBi_2Te_4$ thin films are predominantly synthesized using molecular beam epitaxy (MBE) by co-depositing Mn, Bi, and Te elements on the substrate surfaces.[18, 21-24] Such an approach often leads to the presence of impurity phases such as $Bi_2Te_3$ and $MnTe_2$ clusters.[21, 23, 25] On the other hand, by precisely matching the growth rates of 1 monolayer (ML) MnTe and 1 quintuple layer (QL) $Bi_2Te_3$, Luo et al. recently demonstrated an effective approach for the growth of high quality $MnBi_2Te_4$ films with near-stoichiometric uniformity.[24]

The combination of two binary chalcogenides into one ternary chalcogenide has expanded the scope and applications vdW chalcogenides. $In_2Se_3$ and MnSe are two binary chalcogenides that display unique electronic and magnetic properties, respectively. $In_2Se_3$ is a vdW chalcogenide material with multiple phases and polytypes depending on the coordination of the In atoms and the atomic stacking.[26-30] The different phases and polytypes in $In_2Se_3$ have significantly different properties including in-plane or out-of-plane ferroelectricity and photoelectricity.[31-35] MnSe, in its α-phase (rock-salt cubic), is magnetic semiconductor which has antiferromagnetic ordering with a Néel temperature of ~200 K.[36-40] The wide bandgap of ~3.2 eV in α-MnSe has also made α-MnSe a promising candidate in magnetic-optical and optoelectronic devices.

The combination of the layered $In_2Se_3$ and nonlayered α-MnSe can lead to important applications in spintronics as well as optoelectronics. $Mn_2In_2Se_5$ is a vdW compound formed by two MLs of α-MnSe and 1 QL layer of $In_2Se_3$: it has nine covalently-bonded atomic layers (nonuple layers; NLs) between the vdW gaps with atomic order Se-In-Se-Mn-Se-Mn-Se-In-Se.[41, 42] $Mn_2In_2Se_5$ has displayed strong spin glass characteristics with high level of magnetic frustration due to the competing magnetic interactions in the layered structure.[42] The related compound, $MnIn_2Se_4$, has a septuple (7 atomic layer) layer structure similar to that of $Mn_2In_2Se_5$.[43-45] Yang et al. reported ferromagnetic behavior in $MnIn_2Se_4$ with a perpendicular anisotropy at 2 K and a Curie temperature of ~ 7 K.[44] However, research into the Mn-In-Se family of materials is sparse, with thin films of both materials primarily prepared by exfoliation methods: to the best of our knowledge, only Williams et al. has synthesized bulk $Mn_2In_2Se_5$ crystals so far,[42] and no epitaxial $Mn_2In_2Se_5$ thin film has ever been reported.

In this work, we explore the molecular beam epitaxy (MBE) growths of $Mn_2In_2Se_5$ layers by directly depositing MnSe on top of $In_2Se_3$. We find that in a Se-rich environment, the impinging Mn atoms intercalate into the $In_2Se_3$ layers, transforming $In_2Se_3$ into ternary $Mn_2In_2Se_5$. This transformation takes place over a wide range of substrate temperatures and Se:Mn flux ratio. With an optimized MnSe deposition time, a pure, single-crystalline, (0001)-oriented $Mn_2In_2Se_5$ layer with a controlled thickness can be formed. Increasing the MnSe deposition time results in the formation of uniform and single-crystalline α-MnSe (111) films on top of the $Mn_2In_2Se_5$ layer, despite the typical challenges of growing MnSe in thin film form due to its non-layered crystal structure and the presence of unsaturated dangling bonds on the surface.[36-38, 40] This study thus introduces a new method of synthesizing ternary vdW chalcogenides through direct intercalation of foreign atomic species.

## METHODS

$Al_2O_3$ (0001) wafers (2", Cryscore Optoelectronic Limited) were diced into $1 \times 1$ $cm^2$ pieces and annealed in a box furnace at 1150 °C for 8 hours. The annealed substrates were sequentially cleaned with acetone, isopropyl alcohol, and DI water in ultrasonic bath at room temperature, and were transferred to a UV cleaner to remove organic contaminants. Immediately after the cleaning process, the substrates were loaded into an introductory chamber which has a base pressure of $1 \times 10^{-8}$ Torr and outgassed at 200 °C for 2 hours. Prior to MBE growth, the substrate was transferred into the MBE chamber and thermally annealed in ultrahigh vacuum (UHV) environment for 10 minutes at 800 °C.

All sample growths were performed in a DCA R450 MBE system (instrument details at DOI: 10.60551/gqq8-yj90) with a base pressure of $5 \times 10^{-10}$ Torr. The Bi and Mn fluxes were supplied in individual thermal effusion cells using source material of >5N purity. The In flux was supplied by a dual-filament effusion cell. The Se flux was supplied by a Veeco mark V valved cracker.[46] All fluxes were calibrated using a quartz crystal microbalance (QCM) at the growth position, where the tooling factor for each source material was previously determined by performing physical thickness measurements on calibration samples. The growth steps for the $In_2Se_3$ buffer layer are schematically presented in Figure S1. Similar growth processes for $In_2Se_3$ was previously reported in Ref. [47]. Specifically, a thin (~4 QL) $Bi_2Se_3$ buffer layer was first grown on the $Al_2O_3$ (0001) substrate. The substrate temperature, Se:Bi flux ratio, and film growth rate for the $Bi_2Se_3$ growth are 225 °C, 1.7, and 0.1 Å/s, respectively. Subsequently, $In_2Se_3$ was grown on the $Bi_2Se_3$ buffer at a substrate temperature of 300 °C, Se:In flux ratio of ~1.6, and a growth rate 0.1 Å/s. The sample was then ramped up to 600 °C at 30 °C/min and annealed for 10 minutes to desorb the bottom $Bi_2Se_3$ layer, leaving the free-standing $In_2Se_3$ on the substrate. Comparisons of crystal quality, surface morphology, and Raman spectra for $In_2Se_3$ layers grown with and without the use of $Bi_2Se_3$ underlayer evaporation are displayed in Figure S2. In summary, with the $Bi_2Se_3$ underlayer evaporation, the surface morphology of the $In_2Se_3$ layer is substantially improved with a root-mean-square (RMS) roughness of 0.6 nm (compared to > 1 nm RMS roughness in the sample without $Bi_2Se_3$ evaporation, i.e., direct growth approach), while both the XRD 2θ peak positions and Raman peak shifts in the $A_{1g}^1$, $E_g^2$, and $A_{1g}^2$ phonon modes remain similar between the two samples. The substrate was then ramped down at 20 °C/min to the growth temperature, as described in detail in Table S1, for MnSe growth. The Se flux was supplied throughout the entire process. After the MnSe layer growth, the sample is annealed for 2 minutes in a selenium environment at the growth temperature before being cooled to room temperature at a rate of 50 °C/min.

The structural properties of the samples were studied by x-ray diffraction (XRD) experiments using a Malvern Panalytical Empyrean diffractometer equipped with a copper anode and a PIXcel 3D detector. The incident optics include a divergence slit of 1/8°, a primary mask of 14 mm, and a secondary mask of 6 mm. The surface morphology of samples was examined by Bruker Dimension Icon atomic force microscope (AFM) running at peak force tapping mode with a scan rate of 1 Hz and a lateral resolution of 512 pixels/line. Raman spectra were obtained using a Horiba Scientific LabRAM HR Evolution Confocal Raman Microscope equipped with a 532 nm laser (Oxxius LCX single mode) with an optical power of 35 μW and a Si-array back-illuminated deep-depleted detector. To ensure the best possible spectral resolution, the spectra were recorded using a volume Bragg grating (VBG) notch filter, an 100x objective lens, a confocal aperture of 50 μm, and an 1800 gr/mm diffraction grating. All the *ex-situ* characterization described above were conducted at room temperature.

For cross-sectional scanning transmission electron microscopy (STEM), samples were prepared using a FEI Helios 660 focused ion beam (FIB) system. A thick protective amorphous carbon layer was deposited over the region of interest then Ga+ ions (30kV then stepped down to 1kV to avoid ion beam damage to the sample surface) were used in the FIB to make the samples electron transparent for TEM images. High resolution STEM was performed at 300 kV on a dual spherical aberration-corrected FEI Titan G2 60-300 S/TEM system. All the STEM images were collected by using a high angle annular dark field (HAADF) detector with a collection angle of 50-100 mrad. Energy-dispersive X-ray spectroscopic (EDS) elemental maps of the sample surface were collected by using a SuperX EDS system under STEM mode which has four detectors surrounding the sample.

## RESULTS AND DISCUSSION

### *A. Effect of Substrate Temperature*

We directly deposited MnSe under a wide range of growth conditions on the 5 nm thick $In_2Se_3$ film grown with the $Bi_2Se_3$ underlayer evaporation technique. We first evaluate the effect of substrate temperature ($T_{sub}$) for the growth of MnSe. For these samples, a fixed Se:Mn flux ratio (FR) of 1.7 (Se-rich) is used, along with a fixed layer duration of 26.5 min, which corresponds to a MnSe target thickness of ~ 10 nm. Figure 1 presents the AFM images and RHEED patterns of MnSe grown with $T_{sub}$ from 200°C to 450°C. The bright and streaky RHEED patterns indicate the absence of polycrystalline domains in the layer. All six AFM images show relatively smooth surfaces with triangle-like grain structures, indicating the layer is of three-fold or six-fold symmetry consistent with the $Al_2O_3$ (0001) substrate. With increasing $T_{sub}$, larger grain sizes are clearly observed on the surface, which can be explained by the increase in Mn adatom mobility which leads to a higher degree of layer coalescence and reduced island formation. The improving streaky RHEED patterns along with the reduction in RMS roughness from 3.0 nm to ~1.6 nm among Samples B-F indicate that higher $T_{sub}$ is necessary to obtain a smooth surface. We did not pursue higher substrate temperatures as $In_2Se_3$ roughens above 450°C. RHEED images of Sample D at various times during the deposition are shown in Figure S3 in the Supporting Information.

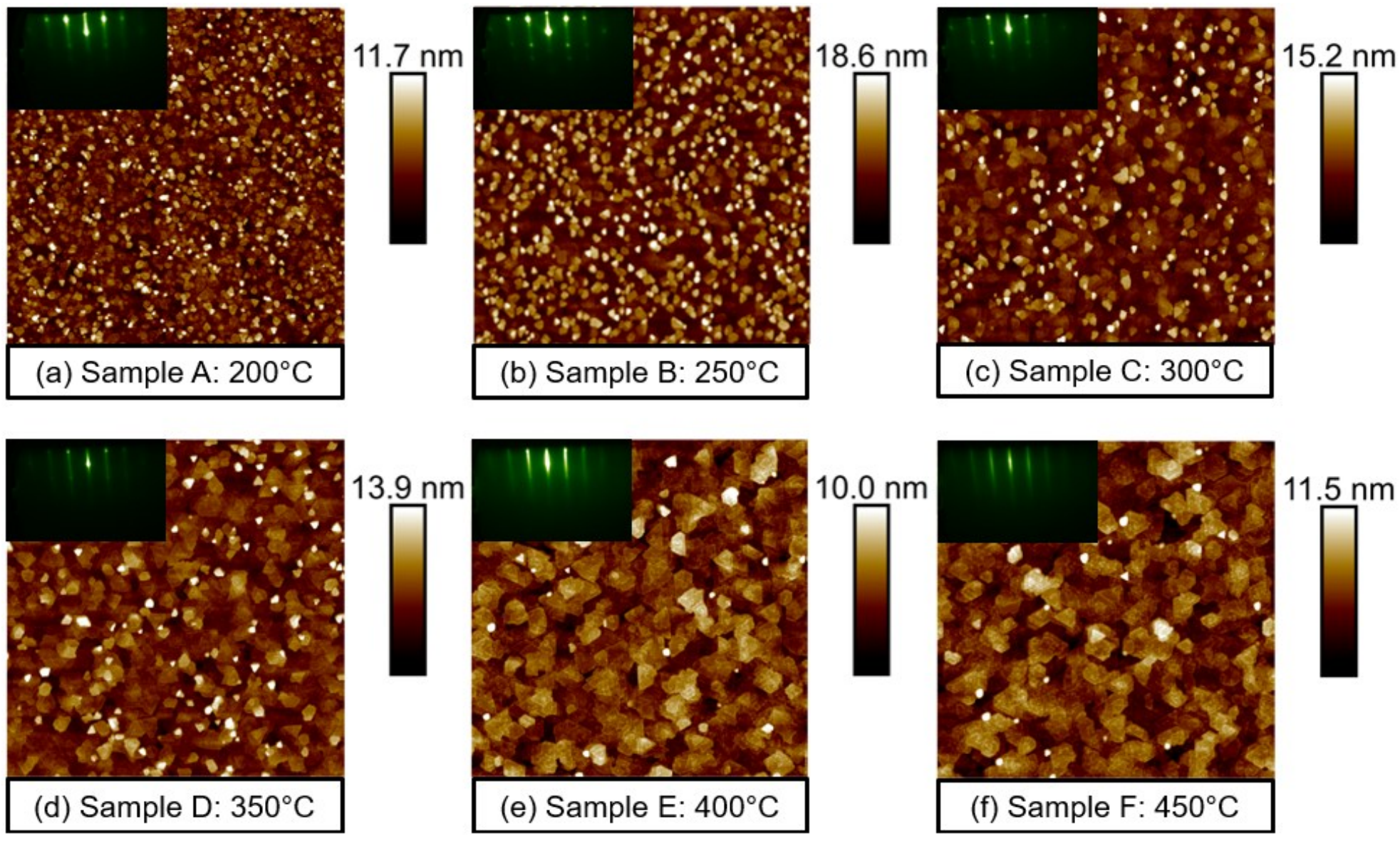

**Figure 1.** (a-f) AFM images of Sample A-F, each grown with $T_{sub}$ ranging from 200°C to 450°C, as indicated by the labels. All images are 2 μm × 2 μm. The RMS roughness for each image is listed in Table S1. Inset shows the corresponding RHEED images taken immediately after the growth at the growth temperature.

Figure 2 shows the XRD 2θ-ω scans of Samples A-F. In all six samples, we see a peak at ~58.8° (dotted line), corresponding to the α-MnSe (222) peak with a lattice constant of ~5.44 Å, which agrees well with the previously reported value. A shoulder at ~29° (dotted line) is also visible, which may correspond to the α-MnSe (111) diffraction peak. However, given the proximity to the stronger peaks at ~ 27°, the exact position of this peak is difficult to determine. Unexpectedly, we do not observe any $In_2Se_3$ $(00m)$ diffraction peaks. Instead, they are replaced by another set of diffraction peaks at ~21.5°, 27.4°, 33.1°, 49.8°, and 56.5° (dashed lines). If we assume that all of these peaks correspond to one different compound, we find that this set of peaks agrees well with the $Mn_2In_2Se_5$ $(00l)$ peaks, where $l = 12, 15, 18, 27, 30$. We note that the $Mn_2In_2Se_5$ $(00\underline{21})$ and $Mn_2In_2Se_5$ (0024) peaks should also be visible, but these two peaks obscured by the $Al_2O_3$ (0002) diffractions. Using the peaks at ~ 21.5° (Samples E and F) and applying the Bragg's law based on the $Mn_2In_2Se_5$ (0012) peak position, the out-of-plane lattice constant of the $Mn_2In_2Se_5$ layers is ~49.5Å, matching well with the results in ref. [42].

The presence of $Mn_2In_2Se_5$ peaks indicate that Mn adatoms, instead of diffusing laterally along the $In_2Se_3$ surface, have also diffused vertically into the $In_2Se_3$ vdW layer, intercalated, and transformed $In_2Se_3$ to $Mn_2In_2Se_5$. We speculate that the smaller atomic size of Mn compared to In and Se promotes intercalation. Furthermore, calculations indicate that $Mn_2In_2Se_5$ has a more negative formation energy compared to $In_2Se_3$ (-0.73 eV/atom in $Mn_2In_2Se_5$ vs. -0.37 eV/atom in $In_2Se_3$),[26, 43] indicating that $Mn_2In_2Se_5$ is more thermodynamically favorable and more stable compared to $In_2Se_3$. Moreover, with increasing $T_{sub}$, a peak shift from 20.7° in Sample A to 21.5° in Sample F can be observed in the $Mn_2In_2Se_5$ (0012) diffraction peak, while the full-width-at-half-maximum (FWHM) of the peak decreases, indicating a change of lattice constant (from 51.5 Å to 49.5 Å) and an increase in the compositional uniformity in the $Mn_2In_2Se_5$ layer. This likely results from the higher percentage of Mn being intercalated and incorporated into the $In_2Se_3$ layer. A similar trend has also been observed in $MnBi_2Te_4$, a vdW compound of which the out of plane peak positions depends heavily on the concentration of Mn in the layer.[21] It is commonly known that the Mn adatom may undergo a variety of processes including interlayer intercalation, desorption, and surface lateral diffusion; the probability of each process depends on the adatom mobility and thus will be strongly affected by the $T_{sub}$. Nonetheless, a wide range of substrate temperature (250-450 °C) can be used for the formation of $Mn_2In_2Se_5$, where the efficiency of Mn intercalation and the uniformity of the layer improves with the higher substrate temperature. We note that we tried co-depositing Mn, In, and Se simultaneously and observed similar XRD patterns to samples grown by intercalation, but the surface is much rougher (see Figure S4).

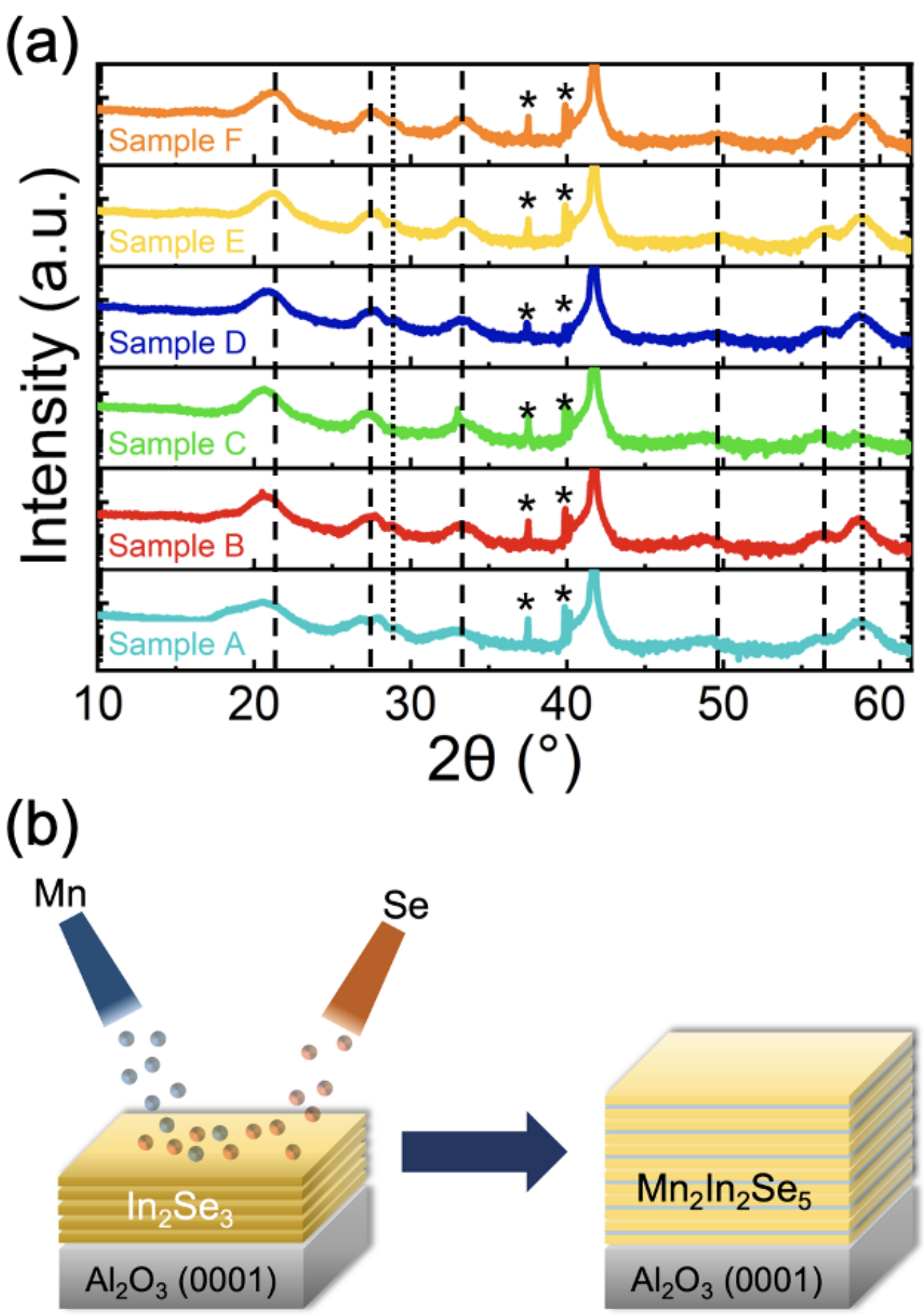

**Figure 2.** (a) XRD 2θ-ω coupled scans of Samples A-F. The peaks marked by a "*" are from the $Al_2O_3$ (0001) substrate. The dotted lines denote (111)-oriented α-MnSe diffraction peaks. The dashed lines denote (0001)-oriented $Mn_2In_2Se_5$ diffraction peaks. (b) Schematic demonstrating the Mn intercalation process that transforms $In_2Se_3$ into $Mn_2In_2Se_5$.

*B. Effect of Se:Mn Flux Ratio*

We further studied the effect of Se:Mn FR on the surface morphology of the samples. In this series, $T_{sub}$ and layer duration are fixed at 400 °C and 26.5 min, respectively, while the Se:Mn FR varies from 1.1 (Sample G) to 3.1 (Sample H). The 2 µm × 2 µm AFM images for these samples are displayed in Figure 3, where the RMS roughness for these three samples are all around 1.5-1.9 nm. Sample E (FR 1.7) is plotted again for comparison. At a lower FR (Sample G), although the RMS roughness remains roughly similar to Sample E, we can see that the grains on the surface are smaller and less dense. This may be caused by the insufficient Se flux and the high Se re-evaporation rate at $T_{sub} = 400$ °C.[48] Increasing the FR to 3.1 (Sample H) results in larger grains with a similar RMS roughness. In the current study, the maximum Se flux was limited to $4.0 \times 10^{13}$ $cm^{-2}/s$ (Sample H), due to the physical limitation of the valve in the Se cracker. However, if we further increased the Se flux, we expect that the surface may become rougher since the diffusion length of the Mn adatoms will be limited by the oversupply of Se, leading to island formation. The XRD 2θ-ω scans of the three samples are presented in Figure S5, where all three samples exhibit peaks consistent with $Mn_2In_2Se_5$ (00$l$) and α-MnSe(111), further confirming a wide Se:Mn FR growth window for the $Mn_2In_2Se_5$/α-MnSe layers.

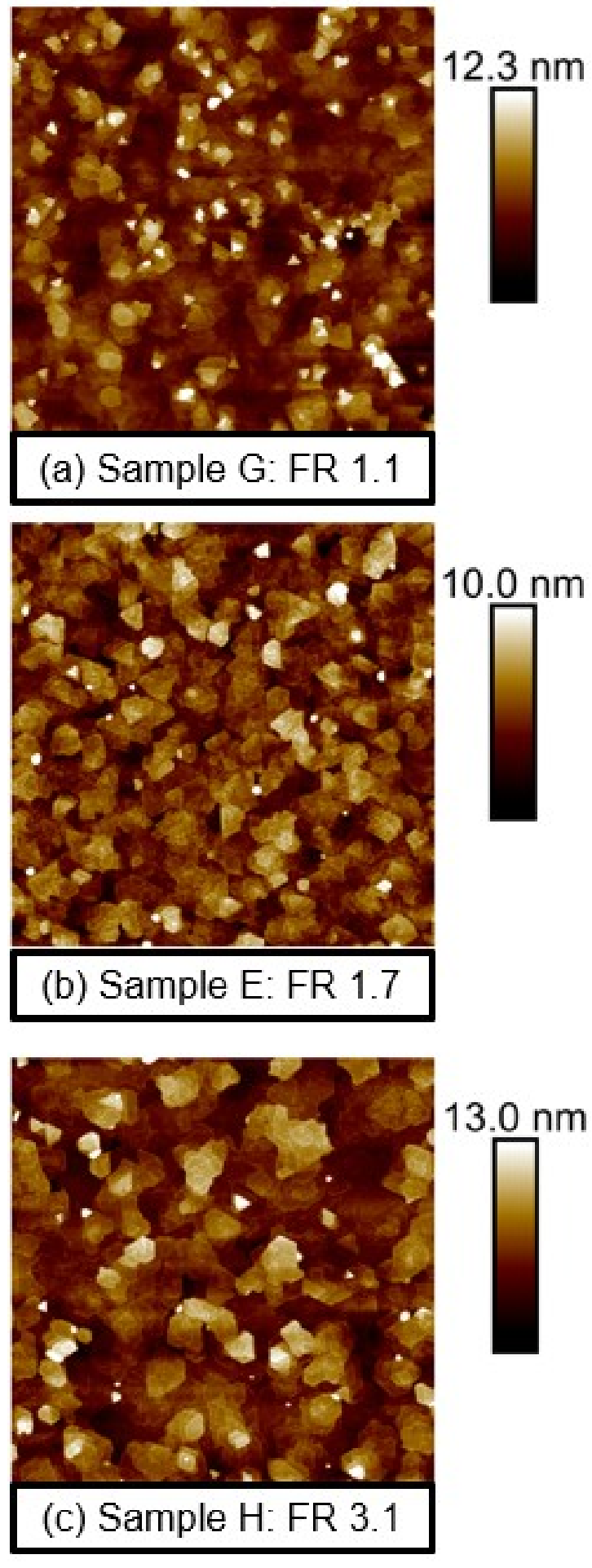

**Figure 3.** AFM images for Samples G (a), E (b, same as **Fig. 1(e)**), and H (c), grown with a FR of 1.1, 1.7, and 3.1, respectively. All images are 2 μm × 2 μm. The RMS roughness for each image is listed in Table S1.

*C. Effect of MnSe Layer Duration*

Using a $T_{sub}$ of 400°C and Se:Mn FR of 1.7, we further carried out growths with different MnSe deposition times. The surface morphology of Samples J-L, grown with a MnSe duration of 2.5 min (Sample J), 10 min (Sample K), and 40 min (Sample L) are shown in Figure S6. All three samples remain relatively smooth with an RMS roughness less than 2 nm. The XRD 2θ-ω scans for these samples, along with Sample E, are presented in Figure 4. In Sample J, we see a broad peak centered near 20°, which can be attributed to a combination of $In_2Se_3$ (006) and $Mn_2In_2Se_5$ (00$\underline{12}$), indicating a mixture of $Mn_2In_2Se_5$ and $In_2Se_3$ in the layer due to an insufficient Mn deposition. Further increasing the layer duration to 10 min (Sample K) results in sharp and clear peaks consistent with (0001)-oriented $Mn_2In_2Se_5$ with no α-MnSe nor $In_2Se_3$ peaks, suggesting that a pure and single-crystalline $Mn_2In_2Se_5$ layer can be formed by optimizing the MnSe duration. Further increases in the MnSe duration (Samples E and L) show a gradual increase in the intensity of the α-MnSe (111) and (222) peaks. We have performed an XRD ω-rocking curve scan of the α-MnSe (222) diffraction peak in Sample L (Figure S7) and found a FWHM of ~0.35°. Compared to typical chalcogenide thin films of similar thickness (~ 8 nm),[39, 49] this value is exceptionally low, and is expected to further decrease with increasing thickness. The increase in the α-MnSe peak intensities also suggest that the α-MnSe layer forms on top of the $Mn_2In_2Se_5$ layer. We thus speculate that the impinging Mn adatoms first intercalate into the $In_2Se_3$ and transform it into $Mn_2In_2Se_5$. After the intercalation is complete, the MnSe layer forms on the $Mn_2In_2Se_5$ surface.

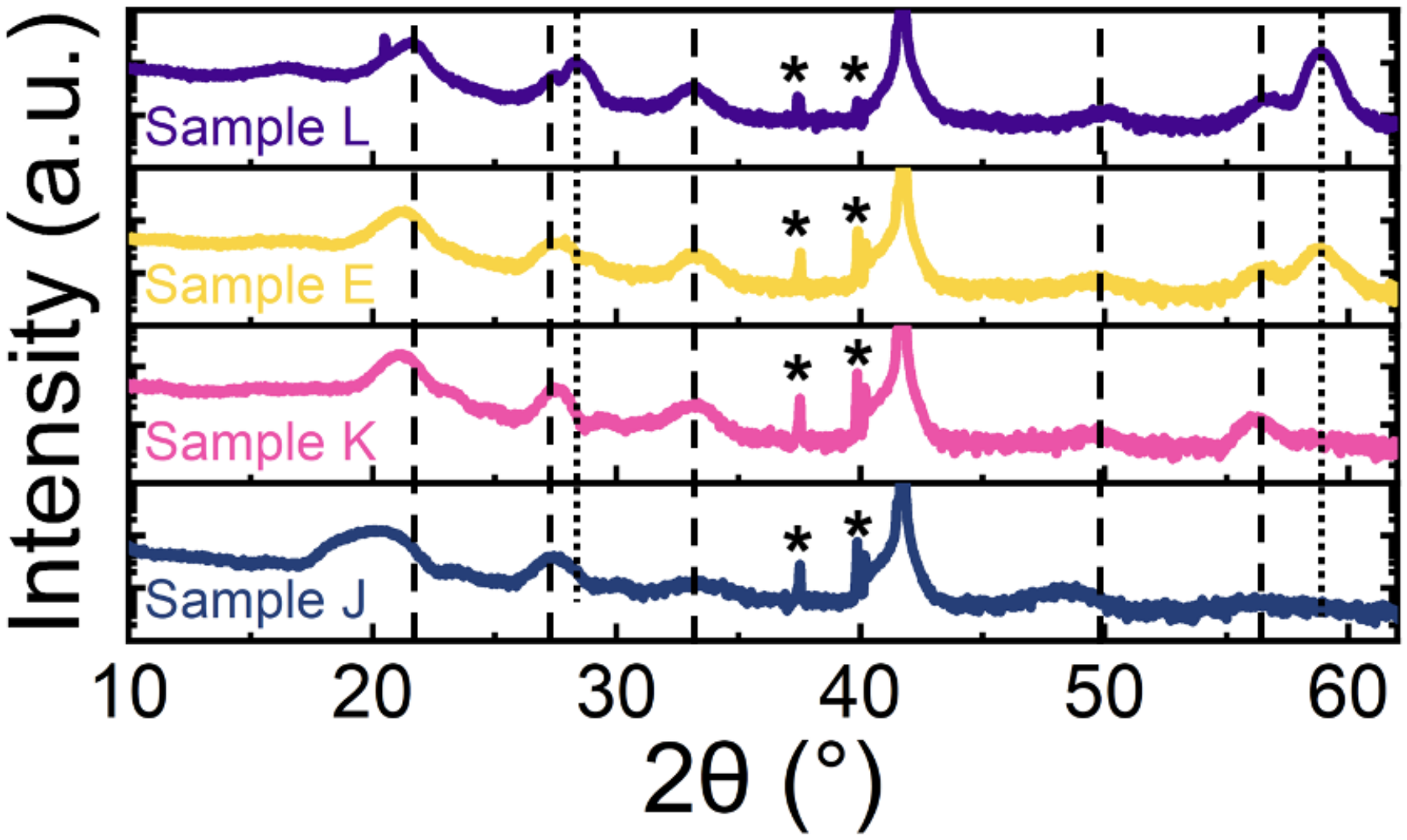

**Figure 4.** XRD 2θ-ω coupled scans of Samples J, K, E, and L, with a MnSe layer duration of 2.5 min, 10 min, 26.5 min, and 40 min, respectively. Sample E is shown again for comparison. The peaks marked by a "*" are from the $Al_2O_3$ (0001) substrate. The dotted lines denote (111)-oriented α-MnSe diffraction peaks. The dashed lines denote (0001)-oriented $Mn_2In_2Se_5$ diffraction peaks.

To confirm this hypothesis, we have carried out STEM experiments on Sample E. Shown in Figure 5 is the cross-sectional HAADF-STEM image along with the In, Se, and Mn energy dispersive spectroscopy (EDS) mapping of the same region. In the HAADF-STEM image, we clearly see two distinct materials. While the In signal is concentrated in the layers immediately adjacent to the $Al_2O_3$ substrate, the Mn signal is detected across the entire thin film but is more concentrated in the top layer. By performing atomic concentration analysis using the K-series electrons of Se, Mn, and In, we have found that the Se:Mn atomic ratio in the top layer is ~1:1, while the Se:Mn:In atomic ratio in the middle layer is close to 2.5:1:1, both of which agree with the stoichiometric ratio of MnSe and $Mn_2In_2Se_5$ suggested in the XRD scans. We can clearly see that the MnSe layer grows on top of the $Mn_2In_2Se_5$ layer, confirming our previous hypothesis. We also observe a thin region of 1-2 nm in thickness with a moderate In signal near the top surface of the sample. This may be caused by the formation of In clusters during the desorption process or by $In_2Se_3$ surface segregation.

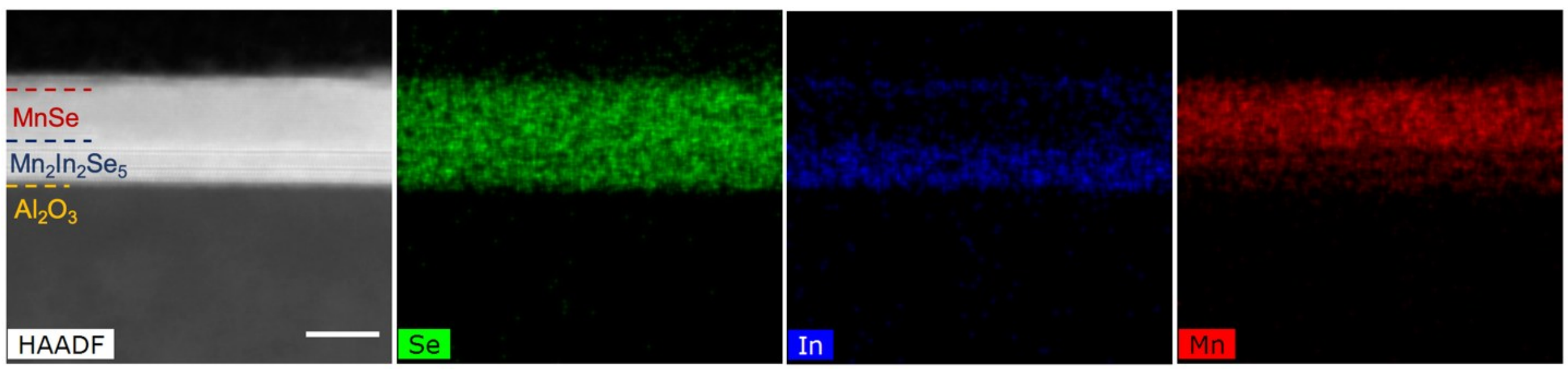

**Figure 5.** Cross-sectional HAADF-STEM image (left) and elemental mapping of Se, In, and Mn in Sample E. The dashed lines in HAADF-STEM image separate the MnSe, $Mn_2In_2Se_5$, and $Al_2O_3$ substrate. The scale bar in HAADF-STEM image is 7 nm.

Multiple regions were scanned throughout the STEM lamella. We find that the thickness of the $Mn_2In_2Se_5$ layer ranges from 5-10 nm, which is ~1-2 unit cells. The thickness fluctuations may result from the thickness nonuniformity in the initial $In_2Se_3$ layer. Shown in Figure 6 (a) is the high resolution HAADF-STEM image of a $Mn_2In_2Se_5$ region. Each $Mn_2In_2Se_5$ layer consists of 9 layers of atoms with the atomic positions mapped in Figure 6 (b). The upper and lower region (where the In and Se atoms are bonded) has a similar structure to 2H β-$In_2Se_3$, while the middle region (where the Mn and Se atoms are bonded) mimics the crystal structure of α-MnSe. Clear gaps separating the layers confirm that this is a vdW structure. It should also be noted that while the majority of $Mn_2In_2Se_5$ NLs are in the $[1\bar{1}00]$ direction, we do see a $Mn_2In_2Se_5$ NL oriented in the $[11\bar{2}0]$ direction, indicative of twin domain formation. These twin domains could originate from the initial $In_2Se_3$ layer, which also shows twin defects.[32] Figure 6 (c-e) shows a HAADF-STEM image of the MnSe layer which is ~5nm in thickness. The MnSe is single-crystalline, with the atomic arrangements confirming its rock-salt crystal structure. A fast Fourier Transform (FFT) confirms that the MnSe is (111)-oriented and has a lattice constant of ~5.4 Å, consistent with theoretical calculations and with the result obtained from the XRD 2θ-ω scan.

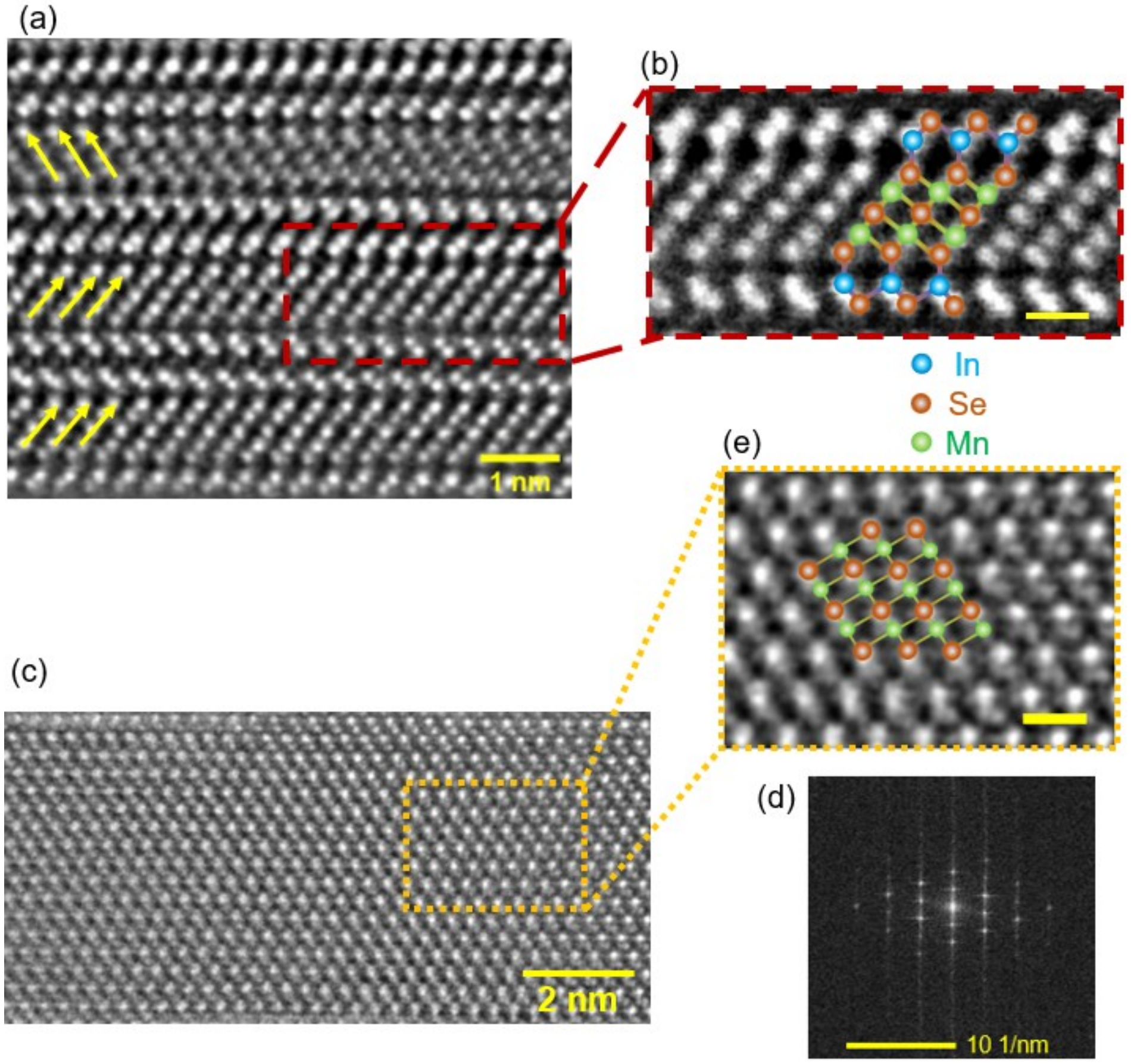

**Figure 6.** (a) HAADF-STEM image of the $Mn_2In_2Se_5$ region. The arrows indicate the orientations of each layer. Arrows pointing up-right indicate $[1\bar{1}00]$ domain orientations, while arrows pointing up-left indicate $[11\bar{2}0]$ domain orientations. (b) High magnification HAADF-STEM image of the boxed region (one $Mn_2In_2Se_5$ NL) highlighted in dashed lines in (a). The scale bar in (b) is 0.5 nm. (c) High resolution HAADF-STEM image showing the MnSe region. (d) FFT analysis of (c), indicative of a single crystalline α-MnSe (111) structure. (e) High magnification HAADF-STEM image of the boxed region in (c) highlighting the MnSe (111) layer. The scale bar in (e) is 0.5 nm.

## SUMMARY

In conclusion, we have reported the molecular beam epitaxy growth and characterization of $Mn_2In_2Se_5$ thin films by intercalating Mn into $In_2Se_3$ under a Se overpressure. Using a 5 nm $In_2Se_3$ buffer layer grown with $Bi_2Se_3$ underlayer evaporation, we explored the growth of MnSe over a wide range of growth parameters, including substrate temperature, Se:Mn flux ratio, and MnSe layer duration. Unexpectedly, after exposing the $In_2Se_3$ layer to Mn and Se fluxes, we found that the impinging Mn atoms intercalate and transform the $In_2Se_3$ into a $Mn_2In_2Se_5$ layer, which has an out-of-plane lattice constant of ~49 Å. We attribute this transformation to the small atomic radius of Mn relative to In and Se and the greater stability of $Mn_2In_2Se_5$ compared to $In_2Se_3$. By varying the duration of the MnSe deposition, single crystalline (0001)-oriented $Mn_2In_2Se_5$ films can be grown. Longer depositions result in the formation of α-MnSe (111) layer with a lattice constant of 5.4 Å on top of the $Mn_2In_2Se_5$ film. STEM imaging further confirms a single crystalline α-MnSe layer on top of a trigonal $Mn_2In_2Se_5$ layer. No other stoichiometries are observed in x-ray or TEM measurements. This work presents a unique technique for the synthesis of $Mn_2In_2Se_5$ epitaxial layers using MBE, a relatively-rare air- and water-stable vdW spin glass material.

ACKNOWLEDGMENT

This research was conducted at the Pennsylvania State University Two-Dimensional Crystal Consortium – Materials Innovation Platform which is supported by NSF cooperative agreement DMR-2039351. The authors appreciate the use of the Penn State Materials Characterization Lab.

AUTHOR INFORMATION

**Corresponding Author**

**Qihua Zhang – qzz5173@psu.edu**

**Stephanie Law – sal6149@psu.edu**

**Data availability statement**

The data sets that support the findings of this study are openly available at: https://doi.org/10.26207/y1ts-cp82.

**Supporting Information**

Schematic of MBE growths of $In_2Se_3$ layer grown with $Bi_2Se_3$ underlayer evaporation; Comparisons of $In_2Se_3$ layer with or without the use of $Bi_2Se_3$ underlayer evaporation; Table listing growth parameters of the MnSe layer in Samples A-L grown on 5-nm $In_2Se_3$ buffer layer; RHEED images of Sample D; XRD 2θ-ω coupled scans and AFM images of $Mn_2In_2Se_5$ layer by simultaneously introducing Mn, In and Se fluxes; XRD 2θ-ω coupled scans for Sample G, E, and H; AFM images for Samples J, K, E, and L; XRD ω-scan of MnSe (222) diffraction peak in Sample L.

For Table of Contents Use Only

# Molecular Beam Epitaxy of $Mn_2In_2Se_5$ van der Waals Layers Using Mn Intercalation

Qihua Zhang, Ke Wang, Wesley Auker, Maria Hilse, Stephanie Law

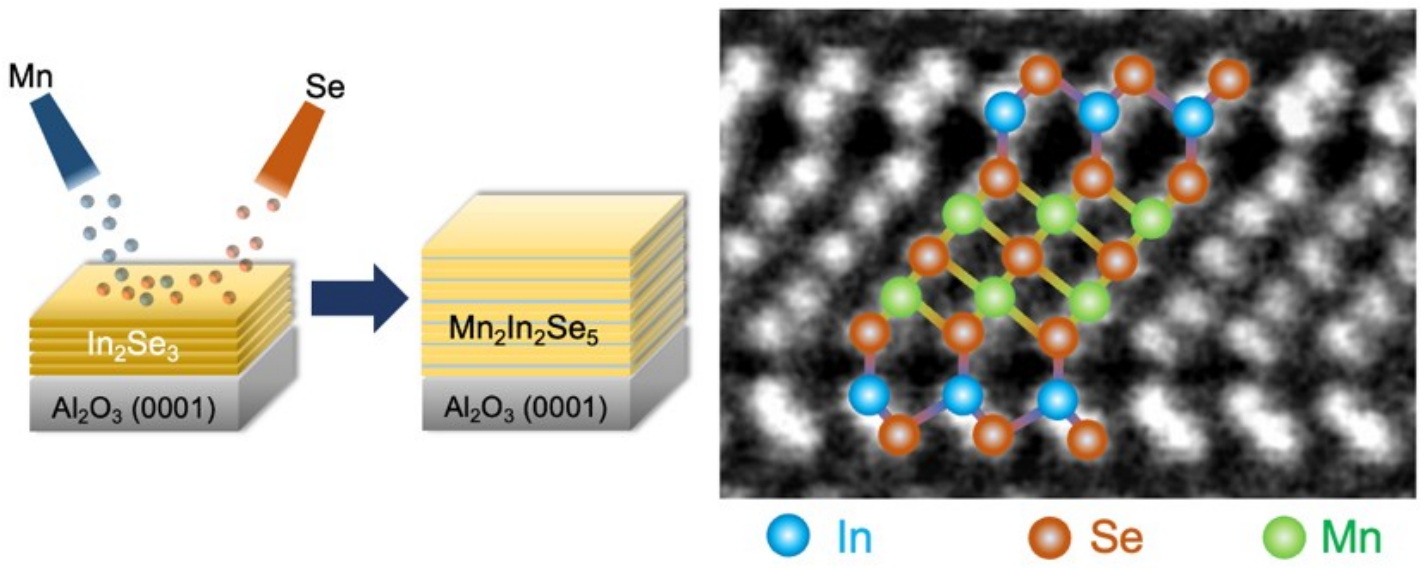

Processes for the synthesis of $Mn_2In_2Se_5$ vdW layers and ball-and-stick model for one nonuple layer of $Mn_2In_2Se_5$.

# Supporting Information

## Molecular Beam Epitaxy of $Mn_2In_2Se_5$ van der Waals Layers Using Mn Intercalation

*Qihua Zhang*[1*], *Ke Wang*[2], *Wesley Auker*[2], *Maria Hilse*[1,2,3], *Stephanie Law*[1,2,3,4]*

[1]Two Dimensional Crystal Consortium Materials Innovation Platform, The Pennsylvania State University, University Park, PA 16802, USA

[2]Materials Research Institute, The Pennsylvania State University, University Park, Pennsylvania 16802 USA

[3]Department of Materials Science and Engineering, The Pennsylvania State University, University Park, Pennsylvania 16802 USA

[4]Institute of Energy and the Environment, The Pennsylvania State University, University Park, PA 16802 USA

CORRESPONDING AUTHORS:

Qihua Zhang – **qzz5173@psu.edu**

Stephanie Law – **sal6149@psu.edu**

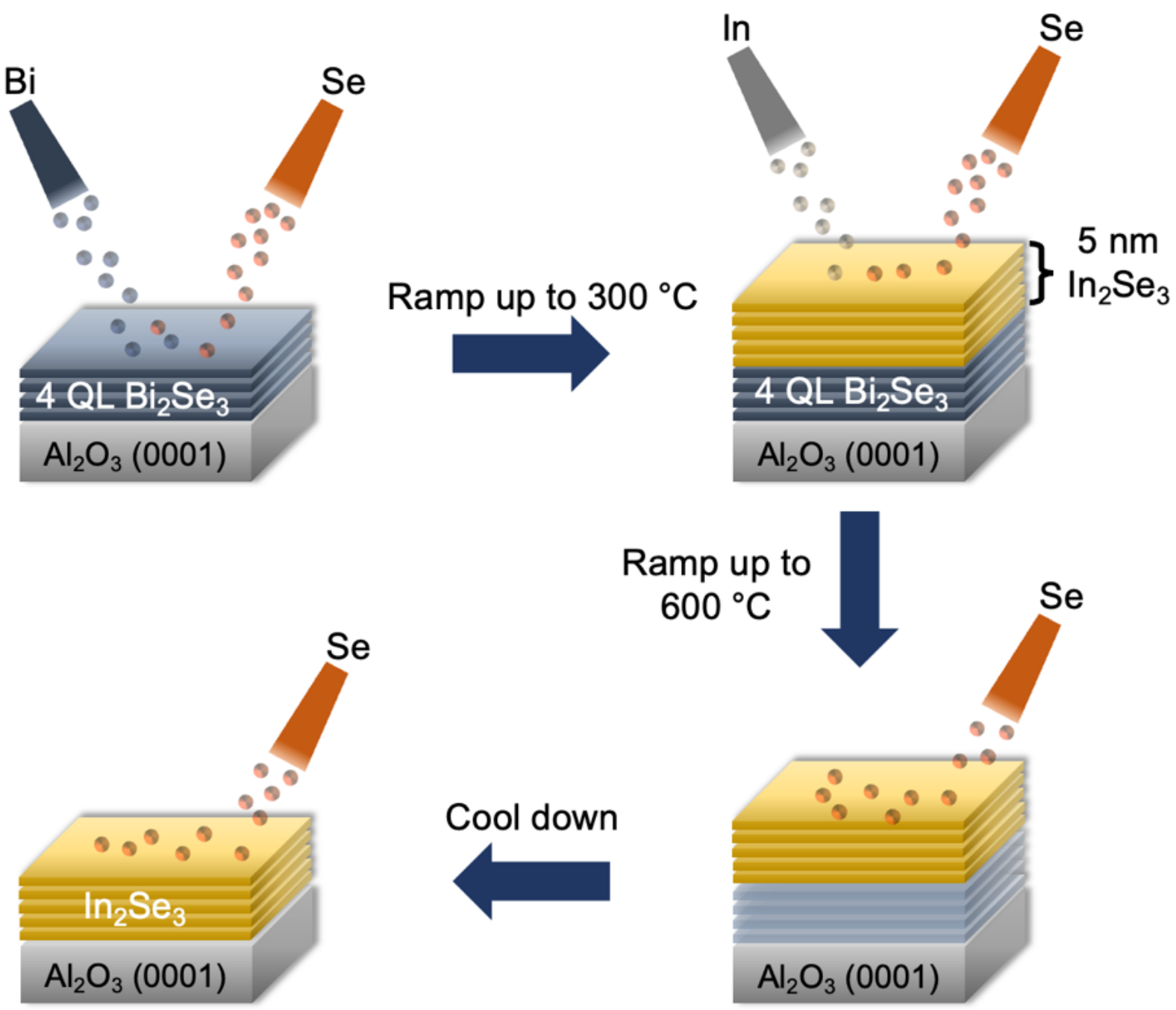

**Figure S1.** Schematic of MBE synthesis process for $In_2Se_3$ layer grown with $Bi_2Se_3$ underlayer evaporation.

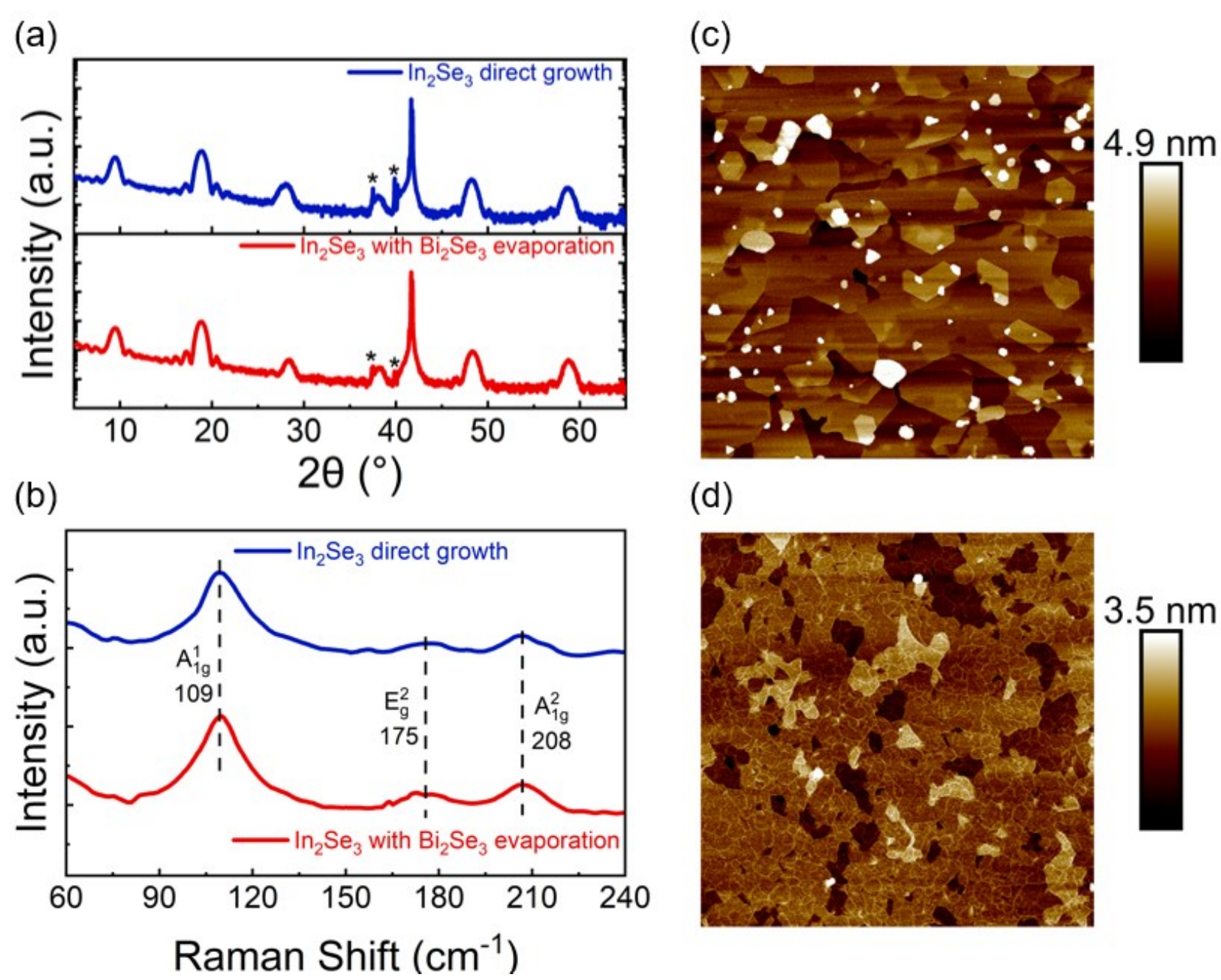

**Figure S2.** Comparisons of surface morphology, crystalline quality, and Raman spectra of $In_2Se_3$ layer with or without the use of $Bi_2Se_3$ underlayer evaporation. Both of the films have a similar thickness of ~10 nm, as examined by x-ray reflectivity (XRR). (a) XRD 2θ-ω coupled scans of a 10-nm $In_2Se_3$ layer grown with $Bi_2Se_3$ underlayer evaporation (red) and with direct growth (blue) on $Al_2O_3$ (0001). All the peaks shown are $In_2Se_3$ $(00m)$ peaks, where $m = 3 \times p$ $(p \in N)$. The peaks marked by a "*" are from the $Al_2O_3$ (0001) substrate. (b) Raman spectra of both samples. (c-d) AFM images with a scan size of 2 μm × 2 μm for the $In_2Se_3$ layer without (c) and with (d) $Bi_2Se_3$ evaporation. The RMS roughness in (c) and (d) is 1.1 nm and 0.5 nm, respectively.

**Table S1.** Growth parameters (Mn flux, Se flux, $T_{sub}$, Se:Mn FR, layer duration) of the MnSe layer in Samples A-L. All samples are grown on 5-nm $In_2Se_3$ buffer layer grown with $Bi_2Se_3$ underlayer evaporation. The RMS roughness of the 2 μm × 2 μm AFM images are also included.

| Sample No. | Mn Flux ($cm^{-2}/s$) | Se Flux ($cm^{-2}/s$) | $T_{sub}$ (°C) | Se:Mn FR | Layer Duration (min) | RMS roughness (nm) |
|---|---|---|---|---|---|---|
| A | $1.3 \times 10^{13}$ | $2.3 \times 10^{13}$ | 200 | 1.7 | 26.5 | 1.7 |
| B | $1.3 \times 10^{13}$ | $2.3 \times 10^{13}$ | 250 | 1.7 | 26.5 | 3.0 |
| C | $1.3 \times 10^{13}$ | $2.3 \times 10^{13}$ | 300 | 1.7 | 26.5 | 2.0 |
| D | $1.3 \times 10^{13}$ | $2.3 \times 10^{13}$ | 350 | 1.7 | 26.5 | 1.9 |
| E | $1.3 \times 10^{13}$ | $2.3 \times 10^{13}$ | 400 | 1.7 | 26.5 | 1.6 |
| F | $1.3 \times 10^{13}$ | $2.3 \times 10^{13}$ | 450 | 1.7 | 26.5 | 1.7 |
| G | $1.3 \times 10^{13}$ | $1.3 \times 10^{13}$ | 400 | 1.1 | 26.5 | 1.9 |
| H | $1.3 \times 10^{13}$ | $4.0 \times 10^{13}$ | 400 | 3.1 | 26.5 | 2.3 |
| J | $1.3 \times 10^{13}$ | $2.3 \times 10^{13}$ | 400 | 1.7 | 2.5 | 0.5 |
| K | $1.3 \times 10^{13}$ | $2.3 \times 10^{13}$ | 400 | 1.7 | 10 | 0.9 |
| L | $1.3 \times 10^{13}$ | $2.3 \times 10^{13}$ | 400 | 1.7 | 40 | 1.5 |

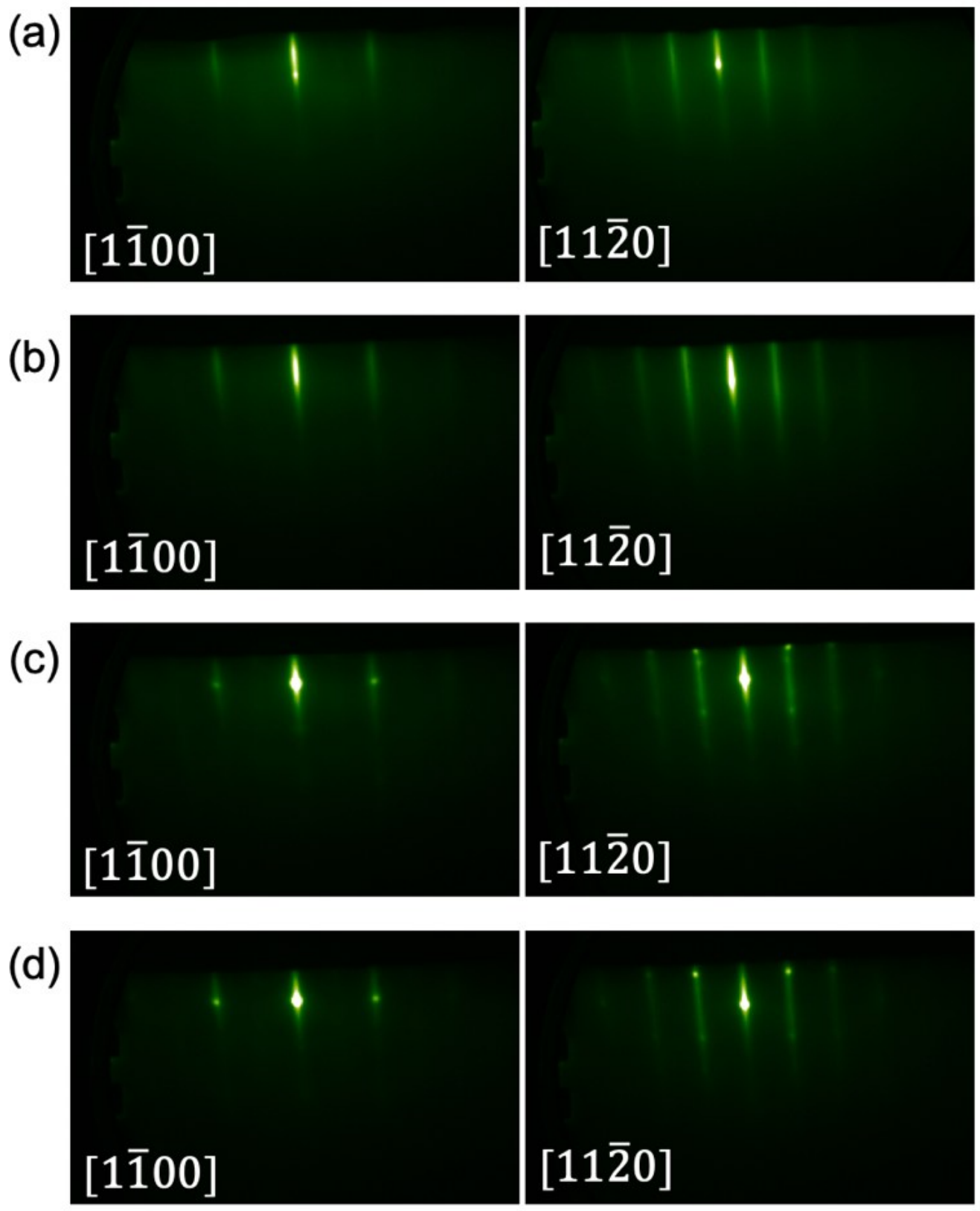

**Figure S3.** RHEED images of Sample D taken along the <11$\bar{2}$0> and <1$\bar{1}$00> direction at the start (a), after 5 min (b), after 10 min (d), and after 26.5 min (d) of Mn+Se exposure. The RHEED images at the start remains similar to that at 5 min. However, after 10 min, the RHEED becomes noticeably spotty, which correlates to the start of MnSe (111) growth.

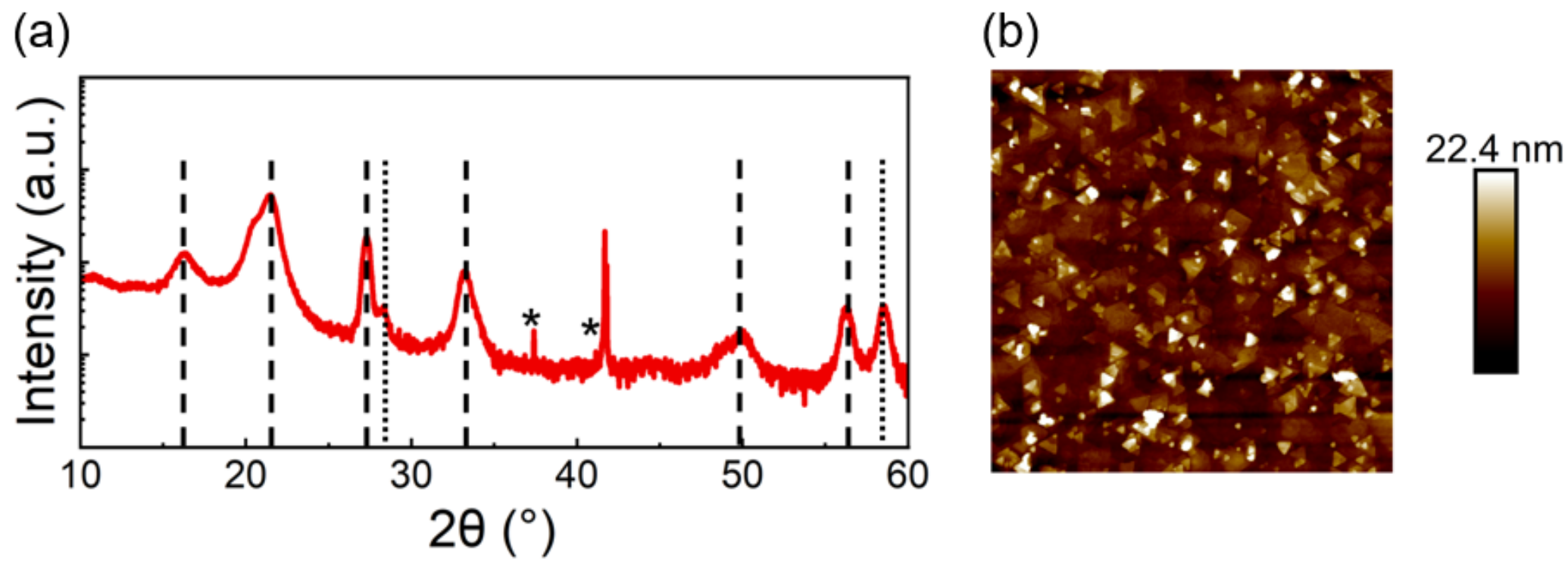

**Figure S4.** (a) XRD 2θ-ω coupled scans of by $Mn_2In_2Se_5$ layer simultaneously introducing Mn, In and Se fluxes. The positions marked by "*" denote the side emissions (Kβ and Lα) of the $Al_2O_3$ (0001) substrate. The dotted lines denote (111)-oriented α-MnSe diffraction peaks. The dashed lines indicate the peak positions related to (0001)-oriented $Mn_2In_2Se_5$ diffraction peaks. (b) 2 μm × 2 μm AFM images of the sample.

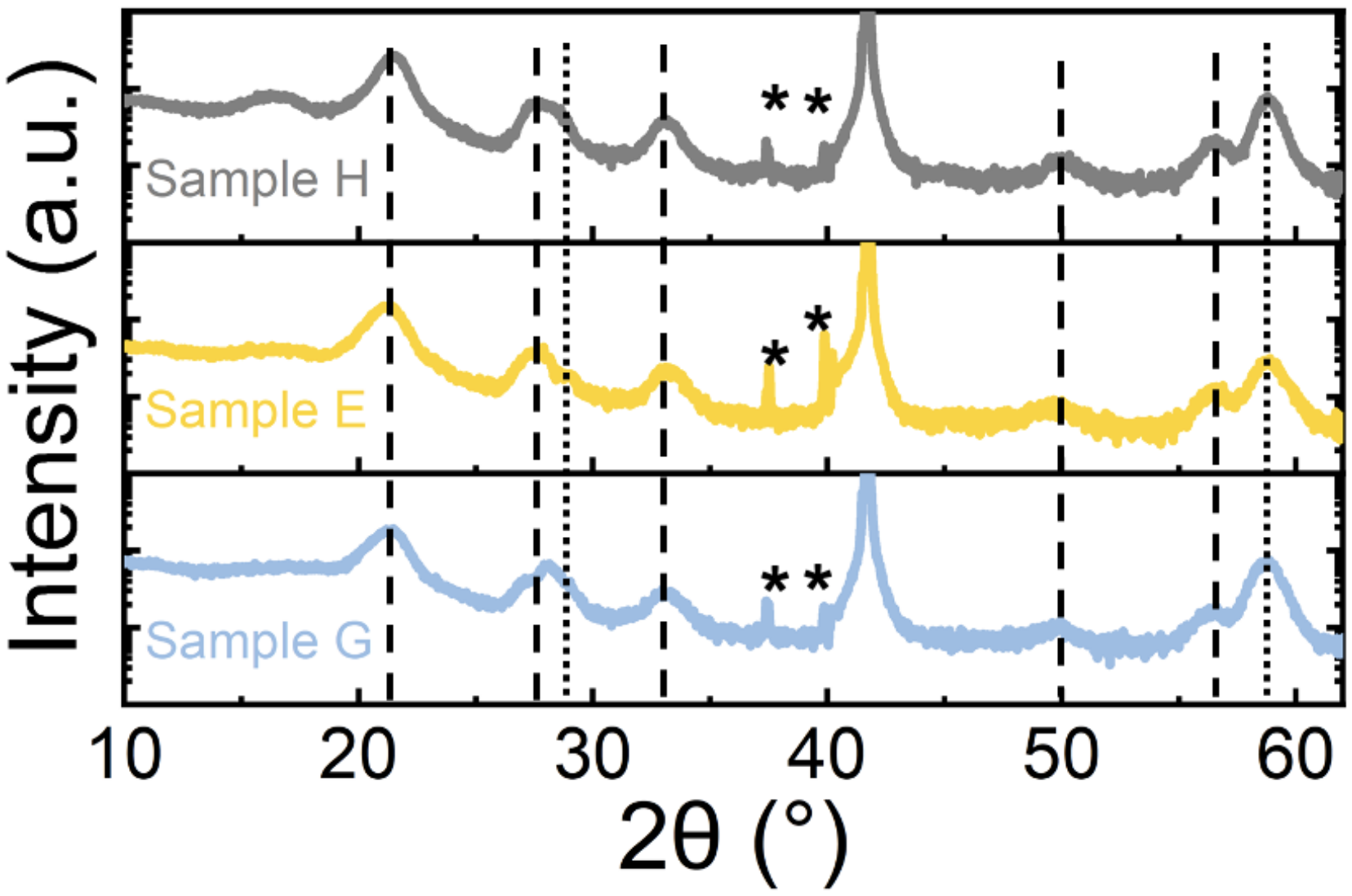

**Figure S5.** XRD 2θ-ω coupled scans for Se/Mn flux ratio study of Samples G (FR = 1.1), E (FR = 1.7), and H (FR = 3.1). The 2θ-ω scan for Sample E (same as second panel from the top in Figure 2) is shown again for comparison. The positions marked by "*" denote the side emissions (Kβ and Lα) of the $Al_2O_3$ (0001) substrate. The dashed lines indicate the peak positions related to (0001)-oriented $Mn_2In_2Se_5$ diffraction peaks. The dotted lines denote (111)-oriented α-MnSe diffraction peaks.

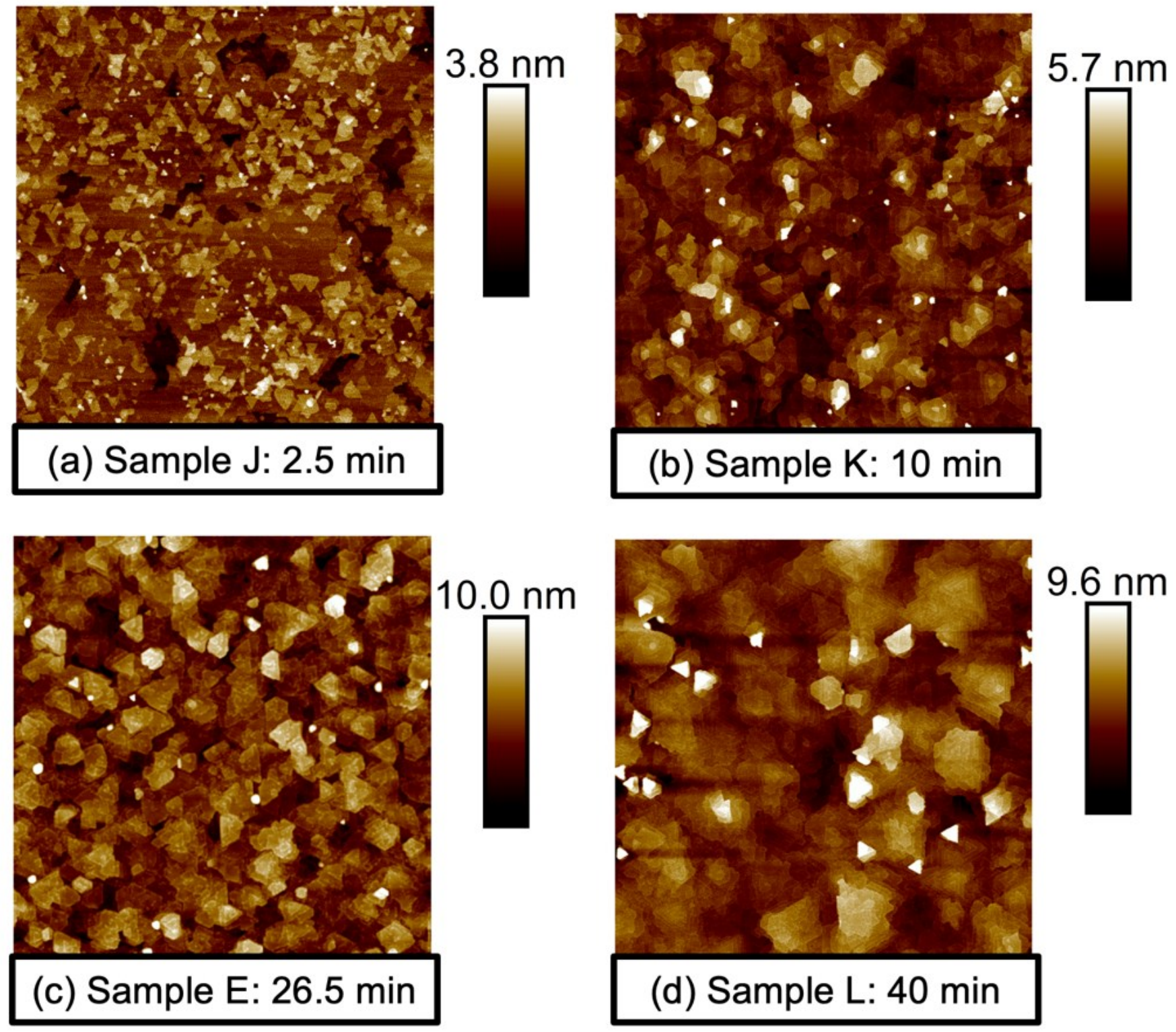

**Figure S6.** AFM images for Samples J (a), K (b), E (c, same as **Fig. 1(e)**), and L (d), each grown with respective MnSe layer duration described in the caption. All images are of 2 µm × 2 µm scan size. The RMS roughness for each image is listed in Table S1.

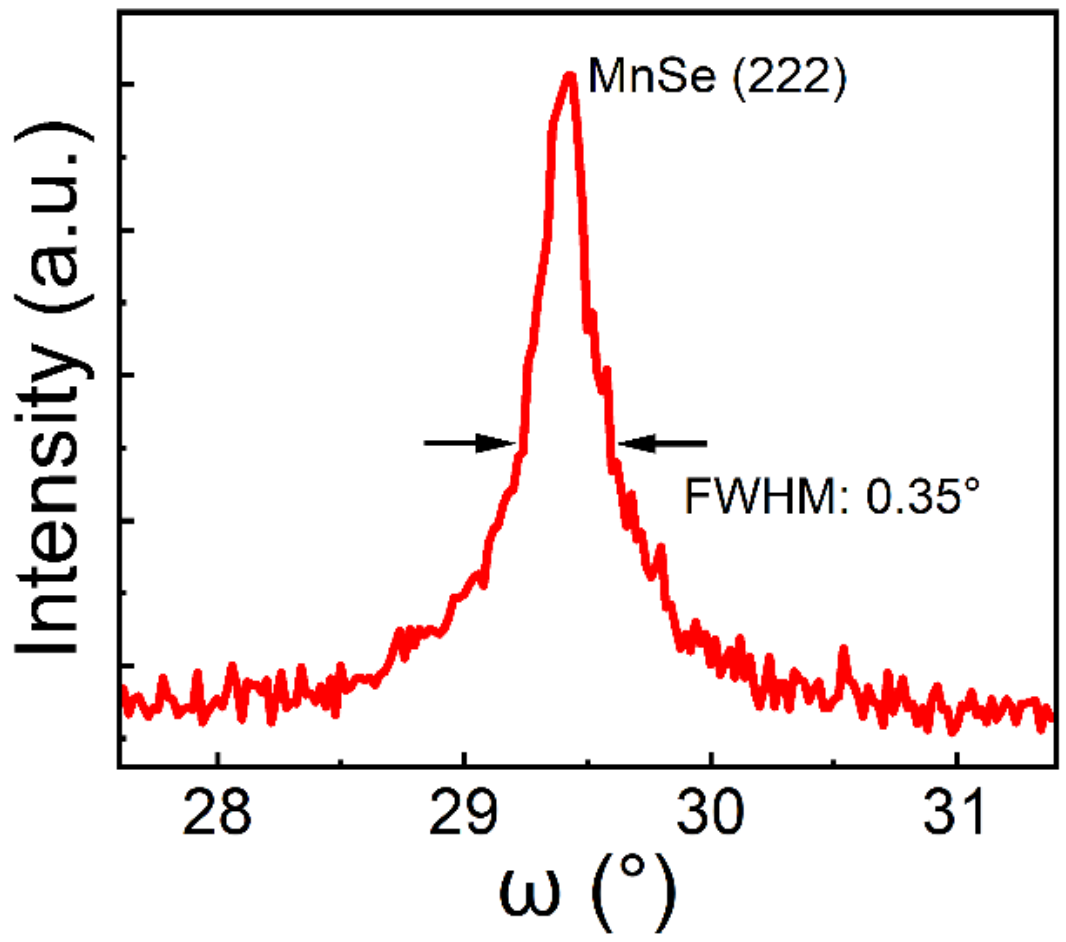

**Figure S6.** XRD ω-scan of MnSe (222) diffraction peak in Sample L.